\documentclass[trackchanges, twocolumn]{aastex701}

\begin{document}

\title{Coalescence Forensics: Weighing the Hosts of Hierarchical Binary Black Hole Mergers}

\author[0000-0001-7197-8899]{Avinash Tiwari} 
\affiliation{Inter-University Centre for Astronomy and Astrophysics, Post Bag 4, Ganeshkhind, Pune-411007, India}
\email[show]{avinash.tiwari@iucaa.in}

\author[0000-0001-5318-1253]{Shasvath J. Kapadia}
\affiliation{Inter-University Centre for Astronomy and Astrophysics, Post Bag 4, Ganeshkhind, Pune-411007, India}
\email[show]{shasvath.kapadia@iucaa.in}  

\begin{abstract}
We present a novel framework to infer the mass of clusters that host hierarchical binary black hole (BBH) mergers detected with gravitational-waves (GWs), {\it on a single event basis}. We show that the requirement that a second-generation (2G) remnant be retained, and subsequently undergo a dynamical encounter, places strong constraints on the mass of the cluster. Using a Plummer model as a readily interpretable baseline, we derive analytic scaling relations between the peak of the inferred host mass posterior, the GW-driven recoil velocity of the remnant, and the parameters that determine the structure of the host. We then perform exact numerical marginalization over thermal and recoil velocities, angles, and cluster structure parameters, to infer the host-mass posterior. We apply our framework to putative hierarchical mergers GW241011 and GW241110, and infer the masses of their hosts on a single-event basis. We find that these are consistent with either heavy globular clusters or nuclear star clusters, with inferred masses spanning $10^{5.7 - 7.7} M_{\odot}$ at $68\%$ confidence depending on the 2G recoil velocity distribution used.

\end{abstract}



\section{Introduction}\label{Sec:Introduction} 
Identifying and modeling the astrophysical origins of compact binary coalescence (CBC) events detected by the LIGO–Virgo–KAGRA (LVK) detector network \citep{aligo, avirgo, KAGRA} is an area of active research \citep{LIGOScientific:2025pvj}. While it is generally accepted that most low-mass CBCs, such as binary neutron stars (BNSs) and neutron star–black hole binaries (NSBHs), form predominantly through isolated binary evolution in the galactic field \citep[see, e.g.][]{Mandel:2021smh}, the formation channels of the LVK’s binary black holes (BBHs) remain uncertain \citep{Mapelli:2021taw}. Identifying the environment of a BBH merger on a single-event basis is particularly challenging for two reasons. First, the current LVK detector network typically provides sky localizations of order tens of square degrees \citep{Ouzriat:2025ben}. Second, the absence of electromagnetic counterparts for BBH mergers precludes unambiguous host-galaxy identification \citep{Chen:2016tys}, and by extension, direct constraints on the location of these events within their host environments.

A number of methods have been proposed to extract intrinsic source-parameter distributions and thereby identify formation channels at a population level \citep{Vitale:2020aaz, ThraneTalbot2019, Talbot:2018cva, ray2023nonparametric}. These approaches have enabled the identification of potential subpopulations \citep{Plunkett:2026pxt, Tong:2025xir, Banagiri:2025dmy, Ray:2024hos, Farah:2026jlc, Vijaykumar:2026zjy, Karathanasis:2022rtr, Afroz:2024fzp, Afroz:2025ikg} within the ever-expanding catalog of LVK BBH mergers \citep{LIGOScientific:2025slb}. Moreover, some studies attempt to infer the provenance of individual BBH events—for example, whether they formed through isolated evolution or were dynamically assembled in dense stellar environments. Such inferences are made by comparing the inferred source parameters—such as component masses and spins—with those expected from BBHs forming in different astrophysical environments \citep[see, e.g, ][for a review of individual event constraints in the context of mergers in dense stellar environments]{Gerosa:2021mno}.

For example, events such as GW190521 \citep{LIGOScientific:2020iuh} and GW231123 \citep{LIGOScientific:2025rsn}—two of the most massive BBH mergers observed by the LVK—feature primary components that lie well within the upper mass gap \citep{Woosley:2021xba}, where black holes formed directly from stellar collapse are not expected. These systems are therefore most naturally explained as products of hierarchical mergers in dense stellar environments \citep{Passenger:2025acb, Li:2025pyo, Paiella:2025qld}. Indeed, GW190521 has been suggested to have occurred in an active galactic nucleus (AGN) disk based on a purported electromagnetic counterpart \citep{Graham:2020gwr}, although this interpretation remains debated \citep{Ashton:2020kyr}. More generally, attempts to identify formation channels for individual BBH events using mass ratio, spin, or eccentricity face substantial degeneracies. For instance, aligned spins can arise both from isolated binary evolution involving a common-envelope phase \citep{Gerosa:2018wbw} and from mergers in AGN disks \citep{Li:2025iux}. Similarly, measurable residual orbital eccentricity may be produced either through dynamical interactions in dense stellar systems \citep{Grishin:2025ofa} or via hierarchical triples in the galactic field \citep{Dorozsmai:2025jlu}. Additional observables, such as line-of-sight acceleration and its higher time derivatives -- particularly with next-generation detectors -- will further sharpen single-event classifications in the future \citep{Tiwari:2024pvb, Tiwari:2023cpa, Tiwari:2025aec, Tiwari:2025qqx, Vijaykumar:2023tjg}.

Recently, the LVK collaboration reported the observation of two BBH mergers -- GW241011\_233834 (hereafter GW241011) and GW241110\_124123 (hereafter GW241110) -- for which there is compelling circumstantial evidence that the more massive component is a second-generation (2G) remnant, although alternative explanations that do not involve dynamical assembly cannot be conclusively ruled out \citep{LIGOScientific:2025brd}. This evidence is primarily based on the asymmetric component masses and the dimensionless spin of the heavier black hole being fully consistent with $\sim 0.7$, both of which are characteristic signatures of hierarchical mergers \citep{Gerosa:2021mno}. Furthermore, the inferred component masses of these systems are consistent with expectations for BBHs formed in dense stellar environments \citep{LIGOScientific:2025brd}, as predicted by large-scale star cluster simulations \citep{rodriguez2022modeling, Antonini:2019ulv}. Under the hypothesis that the primary components of these binaries are 2G remnants of earlier (ancestral) mergers, the recoil velocities imparted during their formation have been estimated \citep{Alvarez:2024dpd, Mahapatra:2021hme, Mahapatra:2024qsy, Mahapatra:2022ngs, Doctor:2021qfn}. These recoil inferences rely on the measured spins of the primary components and are known to depend sensitively on prior assumptions.

In this {\it Letter}, we present a framework to estimate the mass of the host of a hierarchical BBH merger in a dense star cluster on a single-event basis, using recoil velocity posteriors as input. The crux of the method relies on two necessary conditions for a hierarchical merger to occur: (i) the 2G remnant must be retained within the cluster, and (ii) this remnant must subsequently undergo a dynamical encounter. Adopting a Plummer model \citep{Plummer1911}—an analytically tractable approximation for globular clusters (GCs) and non-cuspy nuclear star clusters (NSCs) without central supermassive black holes (SMBHs) — we derive scaling relations connecting retention and encounter probabilities to the cluster mass and structural parameters. We then numerically construct the posterior on the host mass by exactly marginalizing over thermal and recoil velocities, angular configurations, and cluster structure parameters. Applying this framework to GW241011 and GW241110, we find inferred host masses spanning $10^{5.7 - 7.7} M_{\odot}$ depending on the recoil velocity posterior used. This range is suggestive of either massive GCs or NSCs, consistent with theoretical expectations for efficient hierarchical merger environments \citep{Mapelli:2021syv}.

The remainder of this {\it Letter} is organized as follows. In Section~\ref{Sec:Method} we describe the physical framework underlying retention and encounter probabilities and derive analytic scalings using a Plummer model. In Section~\ref{Sec:Results} we present exact numerical marginalization, apply our method to GW241011 and GW241110, and conclude.

\begin{figure*}\label{Fig:Schematic}
    \centering
    \includegraphics[width=1.0\linewidth]{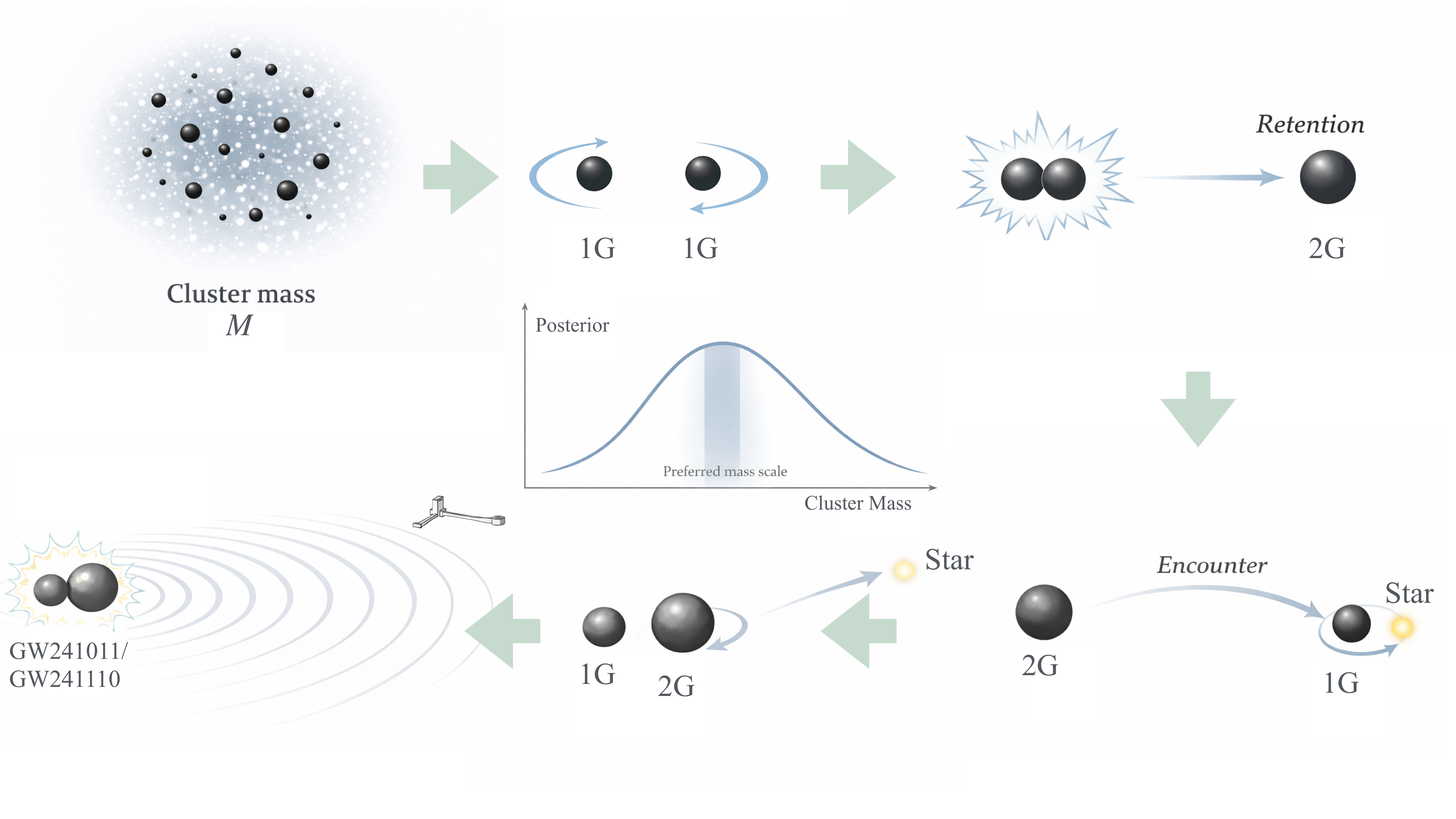}
    \vspace{0cm}
    \caption{Schematic illustration of the sequence of events leading to a hierarchical 1G+2G black-hole merger in a dense stellar cluster. Such a merger requires (i) retention of the 2G remnant following the initial 1G+1G merger, and (ii) a subsequent dynamical encounter between the retained remnant and a stellar system containing at least one 1G BH, shown here for illustration as a 1G+main-sequence-star binary. In this encounter, the lighter main-sequence star is ejected and the 2G remnant forms a binary with the 1G black hole, which subsequently merges. The observed GW signal enables direct inference of the masses and spins of the merging components, and hence a posterior on the recoil velocity of the 2G remnant. Combined with the requirements of retention and encounter, this information yields a preferred mass scale for the host cluster in which the hierarchical merger occurred.}
    \label{fig:placeholder}
\end{figure*}

\section{Method}\label{Sec:Method}
We present a framework to construct the posterior on the mass of a star cluster that hosts a BBH merger, given the hypothesis $\mathcal{H}_{H}$ that the merger is hierarchical. As illustrated in Figure~\ref{Fig:Schematic}, a dynamically assembled 1G+2G merger necessitates that the 2G remnant of the ancestral merger is retained and subsequently undergoes a dynamical encounter. In this analysis, then, we decompose the hierarchical merger hypothesis into two hypotheses: $\mathcal{H}_{H} \equiv \lbrace{\mathcal{H}_{R}, \mathcal{H}_{E} \rbrace}$, where $\mathcal{H}_{R}$ denotes retention and $\mathcal{H}_{E}$ denotes a subsequent encounter. Applying Bayes' theorem, the posterior can be written as:
\begin{equation}\label{Eq:Bayes}
    p(M~|~\mathcal{H}_H, d_{\rm GW}, \vec{p}) = \frac{p(M)P(\mathcal{H}_H~|~M, d_{\rm GW}, \vec{p})}{\mathcal{Z}}
\end{equation}
where $M$ is the mass of the cluster, $d_{\rm GW}$ is the GW data surrounding the detected 1G+2G merger, $\vec{p}$ is a set of additional parameters that define the cluster model, $p(M)$ is the prior, and $\mathcal{Z}$ is the evidence that normalizes the posterior. The likelihood can be written as a product of the retention and encounter probabilities: $P(\mathcal{H}_H~|~M) = P(\mathcal{H}_R~|~M, d_{\rm GW}, \vec{p})P(\mathcal{H}_E~|~\mathcal{H}_R,~M, \vec{p})$. We estimate these by modeling the BBH host as a Plummer sphere, a tractable though approximate analytical model applicable to GCs and non-cuspy NSCs lacking a central SMBH.

As we discuss in subsequent sections, the retention probability is conditioned on the magnitude of the recoil velocity $\vec{v}_r$, and is marginalized with respect to the GW-data-derived posterior on this speed $p(v_r~|~d_{\rm GW})$. On the other hand, as argued later, the encounter probability is independent of recoil velocity. 

\subsection{The Plummer Sphere}

The Plummer sphere \citep{Plummer1911}, originally introduced as a model for GCs, is a spherically symmetric mass distribution characterized by a cored radial density profile with a finite central density \citep{BinneyTremaine}:
\begin{equation}\label{Eq:profile}
    \rho(r) = \frac{3M}{4\pi b^3}\left(1 + \frac{r^2}{b^2} \right)^{-5/2}
\end{equation}
where $b$ denotes the Plummer radius. Physically, this represents a scale radius from the centre of the sphere after which the density drops rapidly. The distribution of radial locations $r$ follows directly from the density profile: $p(r~|~ b) = \frac{4\pi r^2\rho(r)}{M}$.

The gravitational potential and escape velocity are straightforwardly derived to be:
\begin{equation}\label{Eq:potesc}
    \Phi(r) = -\frac{GM}{\sqrt{r^2 + b^2}}, ~~~ v_{\rm esc} = \sqrt{-2\Phi(r)}
\end{equation}
Modeling the star cluster as a collisionless Boltzmann gas with isotropically distributed particle velocities, the standard Plummer velocity distribution is found to be \citep{BinneyTremaine, Aarseth1974}:
\begin{equation}\label{Eq:veldist}
    p\left(v~|~r, \vec{s} \right) = \frac{512}{7\pi}\frac{v^2}{v_{\rm esc}^3}\left(1 - \frac{v^2}{v_{\rm esc}^2}\right)^{7/2}, ~ 0\leq v < v_{\rm _{\rm esc}} 
\end{equation}

It is normalized to ensure that the thermal velocity magnitude $v$ of a particle in the gas never exceeds the escape velocity $v_{\rm esc}(r)$. Here, $r$ is the radial location of the particle, and $\vec{s} = \lbrace{b, M \rbrace}$ are additional structure parameters\footnote{Note that $M \in \vec{s}$, always.}. If the Plummer radius scales with the mass of the cluster: $b = b_0 \left(\frac{M}{M_0}\right)^{\beta}$, as is observed for NSCs and (weakly) for certain GCs, then the structure parameters become $\vec{s} = \lbrace{M, M_0, b_0, \beta \rbrace}$. We refer the reader to Appendix~\ref{app:plum-vel} for a derivation of the Plummer velocity distribution.

\subsection{Retention Probability}

The condition for a 2G remnant with recoil velocity $\vec{v}_r$, to be retained following the ancestral 1G+1G merger is:
\begin{equation}\label{Eq:retcond}
    |\vec{v} + \vec{v}_r| < v_{\rm esc}(r)
\end{equation}
The retention probability is simply the fraction of remnants with recoil speed $v_r$, at location $r$, that satisfy the above inequality. Conditioned on the structure parameters (including cluster mass) and recoil velocity, this probability may be formally written as:
\begin{widetext}
\begin{equation}\label{Eq:retprobfull}
    P\left(\mathcal{H}_R~|~\vec{s}, ~\vec{v}_r \right) = \int_0^{\infty}p\left(r~|~M,~b \right)dr \int_0^{v_{\rm esc}} p\left(v~|~r,~\vec{s} \right) dv \int \Theta\left(v^2_{\rm esc}(r) - |\vec{v} + \vec{v}_r|^2 \right)d\Omega_{\hat{n}} 
\end{equation}
\end{widetext}
where the integrand in the last integral is the Heaviside function, and the integral is performed over isotropically distributed directions $\hat{n}$ of the thermal velocity. Exploiting the isotropy of the Plummer distribution, this triple integral simplifies in two ways. The first (second) is that the marginalization over $\Omega_{\hat{n}}$ ($\vec{v}$) reduces to one over $\cos\theta$ ($|\vec{v}|$), with $\theta$ denoting the angle between the thermal and recoil velocities. We refer the reader to Appendix~\ref{app:angle-avg} for a derivation of the analytical isotropic averaging.

To expose the scaling of the retention probability with the recoil speed $v_r$ and cluster structure parameters $\vec{s}$, it is instructive to approximate the inequality condition to $v^2 + v_r^2 < v_{\rm esc}^2(r)$, with cross-term $2vv_r\cos\theta$ averaging out to zero assuming a simple marginalization of $\theta$ over one cycle. 
As we show in Appendix~\ref{app:meansq-vel}, the squared thermal velocity $\langle v^2 \rangle$ averaged over the Plummer velocity distribution is simply one fourth of the squared escape velocity. The inequality condition for retention then reduces to: $v_r^2 < 3 v_{\rm esc}^2/4$. 

Converting this inequality to one over $r$, we find a retention radius $r_{R} = \sqrt{\frac{9G^2M^2}{4v_r^4} - b^2}$ within which particles with recoil velocity $\vec{v}_r$ are retained, and those outside are ejected from the cluster. The fraction of mass within the retention radius $r_R$ is given by $F(< r_R) = (1 + b^2/r_R^2)^{-3/2}$. Assuming that the number fraction of remnants traces the enclosed mass-fraction, the retention probability is given by:
\begin{equation}
    P(\mathcal{H}_R~|~v_r,~\vec{s}) \approx \left(1 -CM^{2\beta - 2} \right)^{3/2} 
\end{equation}
if $r_R^2 > 0$, and zero otherwise. Here, $C \equiv \frac{4b_0^2v_r^4}{9G^2M_0^{2\beta}}$, and the Plummer radius is assumed to scale with the mass as a power law, as mentioned earlier. The retention probability increases monotonically with $M$, and asymptotes to unity at large $M$. We will use this approximate expression to derive a scaling relation between the mode of the cluster mass posterior, recoil speed, and cluster mass, in Section~\ref{Sec:mode}. Note, however, that all results presented in Section~\ref{Sec:Results} are obtained using the exact numerical marginalization (cf. Eq.~\ref{Eq:retprobfull}).  

\subsection{Encounter Probability}

\begin{table*}
\centering
\footnotesize
\renewcommand{\arraystretch}{1.2}
\begin{tabular}{|c|c|c|c|c|}
\hline
\textbf{Cluster} 
& \textbf{Structure Parameters} 
& \textbf{$v_r$ Posterior} 
& \textbf{GW241011/GW241110 $\log_{10}(M/M_\odot)$} 
& \textbf{Consistent ?} \\
\hline

GC 
& $(b_0, M_0, \beta) = (1\,{\rm pc},10^5\,M_\odot,0)$ 
& Forwards 
& $6.94^{+1.13}_{-0.96}$ / $6.96^{+1.10}_{-0.90}$ 
& Yes \\
\hline

GC 
& $(b_0, M_0, \beta) = (0.5\,{\rm pc},10^5\,M_\odot,0)$ 
& Forwards 
& $6.64^{+1.13}_{-0.95}$ / $6.66^{+1.10}_{-0.90}$ 
& Yes \\
\hline

GC 
& $(b_0, M_0, \beta) = (1\,{\rm pc},10^5\,M_\odot,0)$
& Backwards 
& $8.67^{+1.09}_{-1.07}$ / $8.33^{+1.14}_{-0.92}$
& Tension \\
\hline

GC 
& $(b_0, M_0, \beta) = (0.5\,{\rm pc},10^5\,M_\odot,0)$
& Backwards 
& $8.37^{+1.10}_{-1.07}$ / $8.03^{+1.15}_{-0.93}$
& Tension \\
\hline

NSC 
& $(b_0, M_0, \beta) = (3.3\,{\rm pc},3.6 \times 10^6\,M_\odot,1/3)$
& Backwards 
& $6.13^{+1.05}_{-0.98}$ / $6.26^{+1.40}_{-1.00}$
& Yes \\
\hline

\end{tabular}
\caption{
Single-event inference of the masses of the hosts of putative hierarchical BBH mergers GW241011 and GW241110. Quoted values are medians with 68\% credible intervals.
``Forwards'' and ``Backwards'' refer to two different methods used by the LVK to estimate recoil velocity posteriors under different prior assumptions \citep{LIGOScientific:2025brd}. Posteriors pertaining to rows 1, 3, and 5 are plotted in Figure~\ref{fig:GW_pdfs}. The NSC size-mass scaling in the last row is taken from~\citep{Georgiev2016}.
}
\label{tab:cluster_mass_inference}
\end{table*}

For a hierarchical merger to occur, a retained 2G remnant must subsequently undergo at least one dynamical encounter with another stellar system containing at least one first-generation (1G) black hole. Assuming Poisson statistics, the probability of such an encounter can be written as \citep{Sigurdsson1993}:
\begin{equation}
    P\!\left(\mathcal{H}_E \mid \vec{s}, v, r \right)
    =
    1 - \exp\!\left[-\Lambda(v,r,t)\right],
\end{equation}
where $\Lambda = \Gamma(v,r)\,t$ is the Poisson mean number of encounters over a time interval $t$ following the formation of the 2G remnant. In writing this expression, we assume that, prior to the encounter, the remnant thermalizes onto an orbit whose characteristic radius does not evolve appreciably over the timescale $t$. Under this assumption, the radius 
$r$ at which the encounter occurs is an independent draw from the Plummer radial distribution, and is uncorrelated with the radius at which the 2G remnant was formed \citep{Kritos:2022ggc}.

The local encounter rate $\Gamma(v,r)$ depends on the number density $n(r)$ of potential interaction partners, the interaction cross section $\Sigma$, and the relative velocity $v_{\rm rel}$ between the 2G remnant and the center of mass of the encountered system \citep{BinneyTremaine}:
\begin{equation}\label{Eq:rate}
    \Gamma(v,r) = n(r)\,\Sigma(v_{\rm rel})\,v_{\rm rel}.
\end{equation}
We write the number density as:
\begin{equation}\label{Eq:number-density}
    n(r) = \frac{f_{\rm BH}\,f_{m_{\rm BH}}\,\rho(r)}{\langle m_{\star} \rangle},
\end{equation}
where $f_{\rm BH}$ is the fraction of the cluster mass in systems containing at least one black hole, $f_{m_{\rm BH}}$ is the fraction of such systems that contain the 1G-like component that subsequently participated in the 1G+2G merger, and $\langle m_{\star} \rangle$ is the mean mass of star systems in the cluster.

In the gravitational‐focusing regime relevant for strong dynamical encounters, the interaction cross section is well approximated by:
\begin{equation}\label{Eq:cross-section}
    \Sigma(v_{\rm rel}) = \frac{2\pi G M_{\rm eff} a_{\rm eff}}{v_{\rm rel}^2},
\end{equation}
where $M_{\rm eff}$ is the total mass entering the interaction and $a_{\rm eff}$ is the characteristic distance of closest approach. While the functional dependence of cross-section on the parameters (such as the masses) of the interaction change based on the type of interaction (e.g. binary–single, single–single, and higher‐multiplicity), and nature of the outcome, the $\sigma \propto 1/v_{\rm rel}^2$ scaling is expected to be maintained in hard encounters \citep[see, e.g, ][in the context of exchange, ionisation and resonance pertaining to hard binary-single encounters]{Hut1983, Heggie:1996bs, Fregeau2004, Ginat2021}. Since the 2G remnant is taken to have thermalized prior to the encounter, and assuming that the center-of-mass velocity of the encountered system is an independent draw from the same Plummer velocity distribution, the relative speed may be approximated as $v_{\rm rel} \simeq \sqrt{2}\,v$.

In general, evaluating the encounter probability requires marginalizing over a set of nuisance parameters\footnote{Note that $\vec{N} \subset \vec{p}$ (cf. Eq.~\ref{Eq:Bayes}).},
$\vec{N}=\{f_{\rm BH}, f_{m_{\rm BH}}, \langle m_{\star} \rangle, t, M_{\rm eff}, a_{\rm eff}\}$,
each of which would require additional astrophysical assumptions. However, for a range of astrophysically motivated choices of parameter values, $\Lambda \ll 1$, even for timescales as long as $t \sim \mathcal{O}(1\,{\rm Gyr})$. In this linear regime, the encounter probability reduces to \citep{Fregeau:2006es}:
\begin{equation}
    P(\mathcal{H}_E \mid \mathcal{H}_R, r, \vec{s}) \simeq \Gamma(r)\,t.
\end{equation}
Assuming that the Plummer scale radius scales with cluster mass as $b = b_0 (M/M_0)^{\beta}$, and marginalizing over the Plummer velocity and radial distributions, we find that the encounter probability depends on the cluster mass as:
\begin{equation}
    P(\mathcal{H}_E \mid \mathcal{H}_R, \vec{s}) = A\,M^{(1-5\beta)/2},
\end{equation}
where the prefactor $A=A(\vec{N},b_0)$ absorbs all dependence on the nuisance parameters. In the linear regime, therefore, these parameters affect only the overall normalization of the encounter probability and do not influence its scaling with cluster mass. Moreover, because the remnant is assumed to have thermalized prior to the encounter, the encounter probability carries no explicit dependence on the recoil velocity $v_r$. We refer the reader to Appendix~\ref{app:encounter} for a derivation of the marginalized encounter probability.

\subsection{Mode of the cluster mass posterior}\label{Sec:mode}

\begin{figure*}[t]
    \centering
    \includegraphics[width=0.485\linewidth]{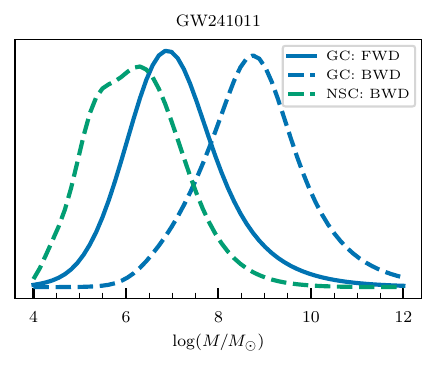}
    \hfill
    \includegraphics[width=0.485\linewidth]{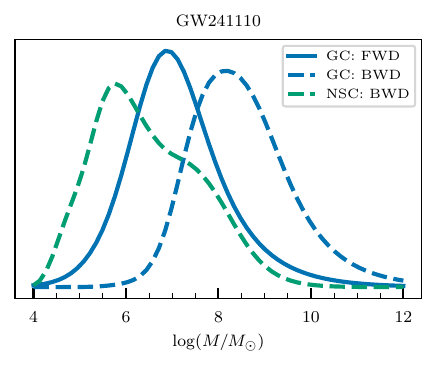}
    \caption{
    Posteriors of inferred cluster mass under different assumptions for
    GW241011 (left) and GW241110 (right). The posteriors correspond to rows 1, 3, and 5 in Table~\ref{tab:cluster_mass_inference}.
    }
    \label{fig:GW_pdfs}
\end{figure*}

To build intuition for the shape of the inferred host-mass posterior, we combine the approximate analytic expression for the retention probability with the encounter probability, assuming a power-law prior on the cluster mass, $p(M) \propto M^{-\alpha}$. Maximizing the resulting posterior yields a closed-form expression for the location of its mode:
\begin{equation}
    M_{\rm max} = \left[\frac{\gamma}{C\left(\gamma + \frac{3}{2}\delta \right)} \right]^{1/\delta},
\end{equation}
where $\gamma \equiv \frac{1}{2} - \frac{5}{2}\beta - \alpha$ and $\delta \equiv 2\beta - 2$. Here, $C$ is the constant defined in Eq.~(9), which encodes the dependence on the recoil velocity and the normalization of the cluster size--mass relation.

Three salient features of the posterior follow immediately. First, the existence of a finite mode requires:

\begin{equation}
    \alpha > \max\!\left(\frac{1}{2} - \frac{5}{2}\beta,\; \frac{\beta}{2} - \frac{5}{2}\right).
\end{equation}

GCs exhibit a weak size--mass scaling \citep[$\beta \simeq 0$, ][]{Gieles2010, Jordan2005, Krumholz2019}, implying that the posterior has no maximum for a uniform prior ($\alpha=0$); in this case, only a lower bound on the cluster mass can be obtained. In contrast, adopting a scale-invariant (log-uniform) prior ($\alpha=1$) yields a unimodal posterior. NSCs, which exhibit a stronger size--mass relation \citep[$\beta \simeq 1/3$, ][]{Georgiev2016, Neumayer2020}, produce a unimodal posterior for $\alpha \gtrsim -1/3$, encompassing both uniform and log-uniform priors.

Second, the location of the posterior mode depends sensitively on the recoil velocity of the 2G remnant. For GC-like scaling, the mode grows as $M_{\rm max} \propto v_r^{2}$, whereas for NSC-like scaling it follows $M_{\rm max} \propto v_r^{3}$. This strong dependence highlights the central role of recoil-velocity inference in enabling meaningful constraints on host-cluster masses from individual GW events.

Third, the posterior mode also depends on the assumed scale radius $b_0$. This dependence is particularly relevant in the context of mass segregation, as more strongly segregated clusters are characterized by smaller effective scale radii. For GCs (NSCs), the mode scales as $M_{\rm max} \propto b_0$ ($M_{\rm max} \propto b_0^{3/2}$). Consequently, more centrally concentrated clusters can support hierarchical mergers at comparatively lower total cluster masses.

\section{Results and Discussion}\label{Sec:Results}

We evaluate the posterior distributions of the host cluster masses for GW241011 and GW241110 under specific assumptions for the cluster structure parameters $\vec{s}$ and adopting a scale-agnostic, log-uniform prior on the cluster mass ($\alpha = 1$). The posteriors are obtained through exact numerical marginalization over angular configurations, the Plummer velocity and radial distributions, and the LVK-provided recoil velocity posteriors $v_r$ for the 2G remnants. We consider both GC-like and NSC–like structural scalings. For the recoil velocity, we employ two alternative LVK posteriors \citep{ligo_scientific_collaboration_2025_17343574}, referred to as the “Forwards” (low-$v_r$) and “Backwards” (high-$v_r$) estimates \citep{LIGOScientific:2025brd}.

The inferred host-mass posteriors for GW241011 and GW241110 are summarized in Table~\ref{tab:cluster_mass_inference} and plotted in Figure~\ref{fig:GW_pdfs}. The results for the two events are similar, which is expected given the overlap between their respective recoil velocity posteriors. We therefore focus our discussion on the general trends revealed by these inferences rather than on event-to-event differences.

Assuming a GC-like size-mass scaling and adopting the low-$v_r$ (“Forwards”) recoil posterior, we find that the inferred host masses are consistent with the upper end of the GC mass spectrum. In contrast, under the same structural assumptions but using the high-$v_r$ recoil posterior, the inferred host masses shift to larger values and are in tension with a GC interpretation at the $68\%$ confidence level, given that typical Galactic globular clusters span masses of $\sim 10^5–10^6\,M_\odot$. When instead adopting an NSC-like size-mass scaling, the same high-$v_r$ posterior yields host-mass inferences fully consistent with low-mass NSCs. We additionally evaluate the posterior for a more centrally concentrated Plummer sphere ($b = 0.5~{\rm pc}$) as an approximate model for mass-segregated GCs, and find that the posterior shifts to smaller values by less than an order of magnitude, and the qualitative results found for un-segregated clusters remain unchanged (see Table~\ref{tab:cluster_mass_inference}). Increasing $b$ by a factor of a few pushes the peak of the inferred mass posterior to larger values by that factor, as expected (cf. Section~\ref{Sec:Method}).

These results numerically confirm the sensitivity of single-event host-mass inference to both the recoil velocity $v_r$ and the assumed size--mass relation, as anticipated in Section~\ref{Sec:Method}. The median recoil velocities inferred using the ``Forwards'' and ``Backwards'' approaches differ by nearly an order of magnitude, which drives a substantial shift in the location of the posterior mode toward higher cluster masses. Conversely, adopting a stronger mass--size scaling ($\beta = 1/3$) suppresses the contribution from the most massive clusters, shifting the posterior toward lower masses. 

Despite the unavoidable sensitivity to these assumptions, the inferred host masses for GW241011 and GW241110 paint a coherent physical picture. Both events are consistent with formation in either high-mass GCs or low-mass NSCs, depending on the recoil velocity posterior adopted. This demonstrates that a dynamics-driven, single-event inference can yield meaningful constraints on the mass of BBH host environments, even in the absence of electromagnetic counterparts or population-level assumptions.

The results presented here rely on several assumptions that merit discussion. First, we model star clusters using a Plummer sphere, which provides an analytically tractable baseline and is commonly adopted for GCs and non-cuspy NSCs whose potentials are not dominated by a central SMBH. While this model does not fully capture the structure of dynamically evolved, mass-segregated clusters or cuspy systems hosting an SMBH, it serves as a transparent starting point for isolating the dominant physical scalings. More realistic cluster models, such as King profiles \citep{king1962structure, king1966structure} calibrated against large-scale star cluster simulations \citep{rodriguez2022modeling, Antonini:2019ulv, Kritos:2022ggc}, represent a natural next step for refining the mass inferences presented here.

Second, we make assumptions about the nature of the dynamical encounter experienced by the 2G remnant. In binary--single interactions, three outcomes are possible---exchange, ionization, and flyby retention. Numerical scattering experiments in the hard-binary regime indicate that the cross sections for these outcomes share a common scaling with relative velocity, $\Sigma \propto v_{\rm rel}^{-2}$ \citep{Heggie:1996bs}. However, this scaling may not hold for soft encounters or for interactions involving higher-order multiples. Incorporating encounter statistics derived directly from $N$-body simulations will allow for a more faithful treatment of encounter outcomes and velocity dependence, which we defer to future work.

Finally, our inference assumes the linear (low-optical-depth) regime for encounter probabilities, which allows the cluster-mass inference to decouple from poorly constrained nuisance parameters such as black hole fractions, encounter timescales, and characteristic closest approach distances. This approximation, adopted in the literature \citep[see, e.g., ][]{Fregeau:2006es}, remains valid across a range of cluster properties relevant to GCs and NSCs. Exploring the boundaries of this regime and relaxing the linearity assumption in denser environments will be an important avenue for future investigation.

\section{Acknowledgments}
We thank Parthapratim Mahapatra for his review and feedback on the manuscript. SJK acknowledges support from ANRF/SERB Grants SRG/2023/000419 and MTR/2023/000086. 

This research has made use of data or software obtained from the Gravitational Wave Open Science Center (gwosc.org), a service of the LIGO Scientific Collaboration, the Virgo Collaboration, and KAGRA. This material is based upon work supported by NSF's LIGO Laboratory which is a major facility fully funded by the National Science Foundation, as well as the Science and Technology Facilities Council (STFC) of the United Kingdom, the Max-Planck-Society (MPS), and the State of Niedersachsen/Germany for support of the construction of Advanced LIGO and construction and operation of the GEO600 detector. Additional support for Advanced LIGO was provided by the Australian Research Council. Virgo is funded, through the European Gravitational Observatory (EGO), by the French Centre National de Recherche Scientifique (CNRS), the Italian Istituto Nazionale di Fisica Nucleare (INFN) and the Dutch Nikhef, with contributions by institutions from Belgium, Germany, Greece, Hungary, Ireland, Japan, Monaco, Poland, Portugal, and Spain. KAGRA is supported by the Ministry of Education, Culture, Sports, Science and Technology (MEXT), Japan Society for the Promotion of Science (JSPS) in Japan; National Research Foundation (NRF) and Ministry of Science and ICT (MSIT) in Korea; Academia Sinica (AS) and National Science and Technology Council (NSTC) in Taiwan.

\newpage
\appendix

\section{The Plummer Velocity Distribution}\label{app:plum-vel}

Following \cite{BinneyTremaine}, we briefly derive the distribution of velocities of the constituents -- assumed to be isotropically distributed -- of a dense star cluster. We model the cluster as a collisionless Boltzmann gas with a spherically symmetric Plummer density profile given by Eq.~\ref{Eq:profile} and Eq.~\ref{Eq:potesc}, respectively. Under these assumptions, Jeans’ theorem implies that the distribution function depends only on the specific energy $E = v^2/2 + \Phi$. Eddington's inversion formula may therefore be applied:
\begin{equation}
    f(E)
=
\frac{1}{\sqrt{8}\pi^2} \frac{1}{M}
\frac{d}{dE}
\int_E^0
\frac{d\rho}{d\Phi}
\frac{d\Phi}{\sqrt{\Phi - E}}.
\end{equation}
to obtain a unique distribution function:
\begin{equation}
    f(E) =  \frac{24\sqrt{2}}{7 \pi^3} \frac{b^2}{(G M)^5}(-E)^{7/2}.
\end{equation}
The velocity distribution, given radius $r$, is obtained to be:
\begin{equation}
    p(\vec{v} \, \vert \, r,\,\vec{s}) = M \frac{f(E)}{\rho(r)} = \frac{128}{7 \pi^2} \frac{1}{v^3_{\rm esc}} \left( 1 - \frac{v^2}{v^2_{\rm esc}} \right)^{7/2}.
\end{equation}
Isotropically averaging over all directions --- $p(v \, \vert \, r,\,\vec{s}) = 4\pi v^2 p(\vec{v} \, \vert \, r,\,\vec{s})$ --- yields Eq.~\ref{Eq:veldist}.
%
%

\section{Isotropic Averaging of the Retention Probability}\label{app:angle-avg}

The condition for a 2G remnant to be retained in the star cluster is presented in Eq.~\ref{Eq:retcond}. Assuming that both the recoil and thermal velocities are isotropically distributed, averaging (cf. Eq.~\ref{Eq:retprobfull}) over all directions reduces to angle averaging over $\cos\theta$, where $\theta$ is the angle between the thermal and recoil velocities. Expanding and rearranging the retention condition, we obtain:
\begin{equation}
    \cos\theta < \mu_{\rm max}, ~\mu_{\rm max} \equiv \frac{v_{\rm esc}^2 - v^2 - v_r^2}{2vv_r}
\end{equation}
Three possibilities immediately emerge: $\mu_{\rm max} \leq -1, ~ -1 < \mu_{\rm max} < 1, ~ \mu_{\rm max} \geq 1$. While the first and third possibilities trivially imply $P_{\rm ret} = 0, 1$, respectively, the second requires an assumption on the distribution of $\cos\theta$, which we take to be isotropic $\cos\theta \sim \mathcal{U}(-1, 1)$. The corresponding distribution is given by $p(\cos\theta) = 1/2$, and the angle averaged retention probability pertaining to this possibility becomes $\int_{-1}^{\mu_{\rm max}}p(\cos\theta)d(\cos\theta) = (\mu_{\rm max} + 1)/2$. We summarize the results of the three possibilities below:
\begin{equation}
P(\mathcal{H}_R \mid r, \vec{s}, \vec{v}, \vec{v}_r)
=
\left\{
\begin{array}{ll}
0, & \mu_{\max} \le -1, \\
\frac{1}2{}(\mu_{\max} + 1), & -1 < \mu_{\max} < 1, \\
1, & \mu_{\max} \ge 1.
\end{array}
\right.
\end{equation}

\section{Mean Square Plummer Velocity}\label{app:meansq-vel}

Starting from the normalized Plummer velocity distribution at fixed radius (cf. Eq~\ref{Eq:veldist}), the local mean squared speed is:
\begin{equation}
\langle v^2 \rangle
=
\int_0^{v_{\rm esc}} v^2\, p\left(v \mid r, \vec{s}\right)\, dv.
\end{equation}
Introducing the dimensionless variable $x \equiv \frac{v}{v_{\rm esc}(r)}$, this becomes:
\begin{equation}
\langle v^2 \rangle
=
v_{\rm esc}^2(r)
\int_0^1 dx\,
\frac{512}{7\pi}
x^4
\left(1 - x^2\right)^{7/2}.
\end{equation}
The remaining integral evaluates to:
\begin{equation}
\int_0^1 x^4 \left(1 - x^2\right)^{7/2} dx
=
\frac{7\pi}{2048},
\end{equation}
yielding:
\begin{equation}
\langle v^2 \rangle
=
\frac{1}{4}\,v_{\rm esc}^2(r).
\end{equation}

\section{Mass-scaling of Encounter Probability}
\label{app:encounter}

We derive the scaling of the encounter probability with cluster mass, marginalized with respect to the Plummer velocity (cf. Eq.~\ref{Eq:veldist}) and radial (cf. Eq.~\ref{Eq:profile}) distributions. From Eqs.~\ref{Eq:rate}, \ref{Eq:cross-section}, the encounter rate for the 2G remnant reduces to:
\begin{equation}
\Gamma(r,v) \propto n(r)\,\frac{1}{v}.
\end{equation}
Marginalizing the encounter rate over the velocity distribution yields (cf. Eqs.~\ref{Eq:potesc}, \ref{Eq:number-density}):
\begin{equation}
\langle \Gamma(r) \rangle_v
\propto
n(r)\int_0^{v_{\rm esc}(r)} \frac{p(v|r)}{v}\,dv \propto
\frac{M^{1/2}b^2}{(r^2+b^2)^{9/4}}.
\end{equation}
Writing $r = b x$ and marginalizing with respect to the Plummer radial distribution:
\begin{equation}
P(\mathcal{H}_E|~\mathcal{H}_R, \vec{s})
\propto
\int_0^\infty
\langle \Gamma(r) \rangle_v \, p(r \, \vert \, b) dr \propto
M^{1/2}\,b^{-5/2}.
\end{equation}
If the Plummer radius scales with mass as $b = b_0\left(\frac{M}{M_0}\right)^{\beta}$:
\begin{equation}
P(\mathcal{H}_E~|~\mathcal{H}_R, \vec{s})
=
A\,M^{(1-5\beta)/2},
\end{equation}
where all numerical factors and nuisance parameters are subsumed into the prefactor $A$.

\bibliography{sample701}{}
\bibliographystyle{aasjournalv7}



\end{document}